\documentclass[preprint,12pt]{elsarticle}

\usepackage{amssymb}

\usepackage{amssymb,amsfonts, amsmath, amsthm,graphicx,caption, subcaption, color, multirow, rotating}
\usepackage[dvipsnames]{xcolor}

\newcommand{\Tau}{\mathcal{T}}
\newcommand{\Sau}{\mathcal{S}}

\newcommand{\dsigma}{~\text{d}\sigma(\mathbf{x})}

\newcommand{\re}{}

\usepackage{comment}

\journal{Mathematical Biosciences}

\date{}

\begin{document}

\begin{frontmatter}



\title{Numerical investigations of the bulk-surface wave pinning model}

\author[GFM,UoS]{Davide Cusseddu}\corref{cor1}
\ead{davide.cusseddu@gmail.com}

\affiliation[GFM]{organization={Grupo de Física-Matematica, Faculdade de Ciências, Universidade de Lisboa},
            addressline={Campo Grande}, 
            city={Lisboa},
            postcode={1749-016}, 
            country={Portugal}}

\author[UBC,UoS]{Anotida Madzvamuse}\corref{cor2}
\ead{am823@math.ubc.ca; a.madzvamuse@sussex.ac.uk}

\affiliation[UBC]{organization={Faculty of Science, Vancouver Campus, Mathematics Department,
121 – 1984 Mathematics Road, Vancouver, B.C. Canada V6T 1Z2}}
\affiliation[UoS]{organization={Department of Mathematics, School of Mathematical and Physical Sciences, University of Sussex},
            addressline={Sussex House}, 
            city={Falmer, Brighton},
            postcode={BN1 9RH}, 
            country={United Kingdom}}

\begin{abstract}
The bulk-surface wave pinning model is a reaction-diffusion system for studying cell polarisation. It is constituted by a surface reaction-diffusion equation, coupled to a bulk diffusion equation with a non-linear boundary condition. Cell polarisation arises as the surface component develops specific patterns. Since proteins diffuse much faster in the cell interior than on the membrane, in the literature, the bulk component is often assumed to be spatially homogeneous. {\re Therefore, the model can be reduced to a single surface equation}. However, {\re in real applications} a spatially non-uniform bulk component might be an important player to take into account. In this paper, we study, through numerical computations, the role of the bulk component and, more specifically, how different bulk diffusion rates might affect the polarisation response. We find that the bulk component is indeed a key factor in {\re determining the surface polarisation response}. Moreover, for certain geometries, it is the spatial heterogeneity of the bulk component that triggers the polarisation response, which might not be possible in a reduced model.
 Understanding how polarisation depends on bulk diffusivity might be crucial when studying models of migrating cells, which are naturally subject to domain deformation.\footnote{This work is dedicated to Professor Edmund J. Crampin (1973 -- 2021).}
\end{abstract}


\end{frontmatter}



\section{Introduction}
Cell polarisation is the result of a very large and intricate network of biochemical and biomechanical processes occurring in the cell, which cause a loss of internal symmetry in its protein distribution and in its shape \cite{Nelson2003}. The biological complexity ranges from and couples different scales, going from molecular interactions to protein reactions and their spatial distributions, to superstructures as filament networks, which support the whole cellular system, ultimately defining the cell shape. 
In order to understand key mechanisms involved in cell polarisation, modellers have being
trying to minimise the number of components and variables under consideration, often reducing their work to purely qualitative descriptions of the polarisation process. 

One of the simplest mathematical models of cell polarisation was originally proposed in \cite{Mori2008} and it is known as the \textit{wave pinning model}. The model is based on the activation-inactivation switching of a representative protein from the GTPase family, and polarisation arises by the appearance of stable regions characterised by high concentrations of the active protein. Originally the model, constituted by a system of two reaction-diffusion equations, was defined on a one-dimensional domain, later extended to two-dimensional domains \citep{Vanderlei2011}.
In \citep{Cusseddu2019} we proposed an extension to a three-dimensional domain $\Omega\subset\mathbf{R}^3$, using a bulk-surface partial differential equation approach, where the active protein was confined to the cell membrane $\Gamma = \partial\Omega$ (a sharp-interface approximation of the cell cortex) and the inactive protein free to move all over the cell $\overline{\Omega}$. We refer to such model as the \textit{bulk-surface wave pinning (BSWP) model} and, for more details on the biological motivations, we refer the interested reader to \citep{Cusseddu2019} and references contained therein. 
In the BSWP model the functions $a (\mathbf{x},t):\Gamma\times[0,T]\to\mathbb{R}$ and 
$b (\mathbf{x},t):\overline\Omega\times[0,T]\to\mathbb{R}$ represent, respectively, the active and the inactive GTPase protein {\re concentrations}, whose evolution is described by the following coupled system of bulk-surface partial differential equations 
\begin{align}
&\frac{\partial b}{\partial t}(\mathbf{x},t)  = D_b \Delta b(\mathbf{x},t), & \mathbf{x}\in\Omega, \label{eq:bulk}\\
&-D_b \frac{\partial b}{\partial \mathbf{n}}(\mathbf{x},t) = f(a,b), & \mathbf{x}\in\Gamma, \label{eq:boundary_condition}\\
&\frac{\partial a}{\partial t}(\mathbf{x},t)  = D_a \Delta_{\Gamma} a(\mathbf{x},t) + f(a,b), & \mathbf{x}\in\Gamma. \label{eq:surface}
\end{align}
Here, $\Delta$ is the classical Laplace operator, $\Delta_\Gamma$ is the Laplace-Beltrami operator, $\mathbf{n}$ is the outward vector to $\Omega$ on $\Gamma$,
$D_b{\re >0}$ and $D_a{\re >0}$ are the bulk and surface diffusion coefficients, respectively. The function $f$ is defined as 
\begin{equation}
    f(a,b) = \left(k_0 + \frac{\gamma a^2}{K^2+a^2}\right)b - \beta a\label{eq:f(a,b)}
\end{equation} 
and indicates the {\re flux, represented by the} reaction between $a(\mathbf{x},t)$ and $b(\mathbf{x},t)$, which takes place at the boundary {\re $\Gamma$} of $\Omega$. The constant parameters $k_0>0$ and $\beta>0$ represent, respectively, the basal activation and inactivation rates, while $\gamma>0$ weights a nonlinear term  describing a positive feedback loop in activation, in the form of a Hill function. At saturation of $a$, the extent of this term tends to $\gamma b$, while $K$ represents the half-activation concentration of $a$.
All the parameter values used throughout this text are reported in \ref{app:numerical}. In what follows, we will consider the above system as a non-dimensional re-scaled version of the BSWP model proposed in \citep{Cusseddu2019}. 
Three key properties of the BSWP model are:
\begin{enumerate}
\item Temporal conservation of total mass, meaning $\frac{d}{dt}\left(\int_\Omega b + \int_\Gamma a\right) = 0$; 
\item $f(a,b)$ admits three zeros $a_1(b)<a_2(b)<a_3(b)$ for $b$ ranging in a certain interval $[b_1, b_2]$; 
\item Difference in the diffusion coefficients $D_a\ll D_b$, which relies on the fact that protein  diffusion on the cell membrane occurs much slower that within the cytosol. 
\end{enumerate}
The {\re \textit{wave pinning model} by Mori et al. \citep{Mori2008}} shares the same properties, but the components $a$ and $b$ diffuse and react on the same spatial domain $\Omega\subset\mathbb{R}^d$, with $d=1$ or 2, and are subject to zero-flux boundary conditions. For our settings, the surface reaction-diffusion equation does not have boundary conditions since it is posed on a closed manifold where $\partial\Gamma=\emptyset$, i.e. it is empty. 
Initial conditions, for both $a$ and $b$, are prescribed in  equations \eqref{eq:ic_a} and \eqref{eq:ic_b} that follow below in our exposition.
For the analysis on the well-posedness of bulk-surface reaction-diffusion systems, we refer the interested reader to the work by Sharma and Morgan \citep{sharma2016global}.

In \citep{Cusseddu2019} we described how the BSWP model might generate cell polarisation. For convenience, we briefly describe it here. In a given parameter region, equation \eqref{eq:surface} is a bistable surface reaction-diffusion equation. This can cause initial perturbations in $a$ to evolve in propagating fronts. The speed of such propagation strictly depends on the bulk component $b$, which, due to conservation of total mass,  gets overall depleted as $a$ expands on $\Gamma$. Eventually, $b$ reaches a critical value $b^*$, causing a pinning of the propagating fronts of $a$. As a consequence, the system approaches an {\it apparent} stationary state, in which 
$\Gamma$ has regions where $a$ has approximately  high values of concentration (denoted here as $a_{high}$) and regions where $a$ has approximately  low values of concentration (denoted here as $a_{low}$), with $a_{high}>a_{low}$ (see also the work done by Brauns et al. on reaction-diffusion systems with conservation of total mass \citep{Brauns2020}). In general, we say that polarisation occurs when $\Gamma$ shows at least one region where $a\approx a_{high}$. 
We will refer to such a region as the \textit{polarisation patch}. 

Extending {\re the original wave pinning model} in \citep{Mori2008} from one to two and three dimensions results in more complex system dynamics. Jilkine \citep{JilkinePhD} and Vanderlei et al. \citep{Vanderlei2011} initially studied a two-dimensional version of the  model, with $a$ and $b$ defined on the same domain $\Omega$. Despite sharing initial propagation and pinning dynamics, in multi-dimensional domains the pinned front is subject to a slow motion across the domain. 
Same dynamics characterise also the bulk-surface wave pinning model \eqref{eq:bulk}-\eqref{eq:f(a,b)} for which, in \citep{Cusseddu2019}, we showed the slow motion of a single polarisation patch towards areas of $\Gamma$ with higher curvature.
Often this shifting occurs on such a long time scale to the extent that it is often neglected in  biological studies or simply that this phenomena is not apparent to trigger biological interest. However it surely triggers a mathematical curiosity for identifying parameter regions and geometries $\Omega$ for which such behaviour might be of biological interest. 

A second characteristic of the wave pinning model, common to all of its different versions, is a sort of competition between different polarisation patches. Indeed, it has been shown that stationary solutions with multiple polarisation patches are not stable  {\re neither in the original wave pinning model by Mori et al. \citep{Mori2008} {\re nor} in its two-dimensional extension presented in \citep{JilkinePhD, Chiou2018, Brauns2021-wavelength}}. A first example of competition in the bulk-surface wave pinning model \eqref{eq:bulk}-\eqref{eq:f(a,b)} was shown in \citep{Cusseddu2019}. {\re Interestingly, in a very recent work, Miller et al. \citep{Miller2022} studied a reduced version of the same model, showing that two patches may instead coexist on non-convex surfaces $\Gamma$.} 

In this work, through numerical investigations, we try to advance current-state-of-the-art understanding of these two characteristics.
To the best of our knowledge, we analyse such dynamics for the first time, taking advantage of the bulk-surface finite element method for solving the system \eqref{eq:bulk}-\eqref{eq:f(a,b)} on different geometries \citep{Cusseddu2019, MADZVAMUSE20169}. {\re Other numerical approaches to similar problems might be found in the literature. As an example we mention the bulk-surface Virtual Element Method, which  is a promising numerical method recently applied by Frittelli et al. for solving the bulk-surface wave pinning model on a two-dimensional spatial domain \citep{Frittelli2021, frittelli2021virtual}.}

{\re We organise this work as follows. In Section \ref{sec:model_reduction} we introduce a reduced version of the BSWP model and discuss about curvature-driven polarisation. In the following, Section \ref{sec:curvature-driven polarisation}, we investigate the influence of the bulk component $b$ in the long time patterning of the solution $a$ on a simple domain, by comparing the model solution for different bulk diffusion coefficients $D_b$.} In Section \ref{sec:competition}, 
we investigate multi-patch competition both on three-dimensional and two-dimensional geometries. In this last case, the surface naturally reduces to a closed curve.
In Section \ref{sec:bulk induces polarisation}, we show an example in which bulk heterogeneity is the main driver of polarisation on the surface. Finally, we conclude our study by summarising our main findings in Section \ref{sec:conclusion}.

\section{Bulk diffusion, model reduction and curvature-driven polarisation}\label{sec:model_reduction}
The bulk component $b$ is the fuel for the propagation and pinning of the activated patch on the surface. However, given the big difference in diffusion coefficients between bulk and surface, $b$ is often considered to be spatially homogeneous. For instance, 
in the analysis of the BSWP model on a disk, in \citep{Cusseddu2019},  we showed that the solution $b$, to a certain approximation, was spatially uniform. Diegmiller et al. \cite{Diegmiller2018} showed that reducing the bulk-surface model to a single reaction-diffusion equation on a sphere still provides an accurate description of the polarisation dynamics. Such reduction results from considering the limit $D_b \to \infty$. Given the conservation of total mass $M(t) = \int_\Omega b + \int_\Gamma a = M_0$, for all $t>0$ and assuming $b$ to be spatially homogeneous at all times, we have $b = \frac{1}{|\Omega|}\left( M_0 - \int_\Gamma a \right)$, which depends only on time. Exploiting this assumption, one then obtains the following reduced surface reaction-diffusion equation
\begin{align}
    \frac{\partial a}{\partial t} = D_a \Delta_\Gamma a +\frac{1}{|\Omega|}\left( M_0 - \int_\Gamma a \dsigma \right)
    \left(
    k_0 + \frac{\gamma a^2}{K^2 + a^2}
    \right) 
    - \beta a, \; \mathbf{x}\in\Gamma. \label{eq:reducedmodel}
\end{align}
{\re We couple} the BSWP model \eqref{eq:bulk}-\eqref{eq:f(a,b)} and the above reduced model \eqref{eq:reducedmodel} to the following Gaussian initial condition for $a$
\begin{equation}\label{eq:ic_a}
    a(\mathbf{x},0) = a_{p,0}\exp{\left\{-\frac{(x-x_0)^2}{\sigma_{x,0}^2} - \frac{(y-y_0)^2}{\sigma_{y,0}^2} - \frac{(z-z_0)^2}{\sigma_{z,0}^2}\right\}}
\end{equation}
and, when solving the BSWP model \eqref{eq:bulk}-\eqref{eq:f(a,b)}, the initial condition for $b$ is prescribed as 
\begin{equation}\label{eq:ic_b}
b(\mathbf{x},0)=\frac{1}{|\Omega|}\left( M_0 - \int_\Gamma a \dsigma \right).
\end{equation}
{\re The initial condition \eqref{eq:ic_a} can be seen as a localised perturbation of the surface protein distribution. However, it is important to remark that this is just one of many possible choices. For instance, polarisation may arise even when the initial datum for both $a$ and $b$ is spatially uniform due to the geometry of the domain \citep{Cusseddu2019} or when $a$ is randomly perturbed across the whole surface $\Gamma$ \citep{Miller2022}.}

The initial peak in $a(\mathbf{x},t)$ may propagate by developing travelling fronts, extending into a \textit{mesa}, i.e. a high plateau region with $a\approx a_{high}$, whereas, in the rest of the surface $\Gamma$, $a\approx a_{low}$.
On a flat surface, the normal speed of a travelling wave in excitable media is given by $v = c{\re(b)} - \kappa D_a$, with $c$ being a non-decreasing function of $b$ and $\kappa$ denotes the curvature of the patch interface \citep{tyson1988singular}. On a generic surface $\Gamma$, $\kappa$ is the geodesic curvature of the front line \citep{bialecki2020traveling}. Therefore, in this last case, also the geometry of the surface plays an essential role in the propagation of $a(\mathbf{x},t)$. 

It was suggested that the motion of the polarised patch across the domain corresponds to a minimisation of the perimeter of the interface separating the states $a_{high}$ and $a_{low}$, under the constrain of constant polarised mass \citep{JilkinePhD}. 
Recently, Singh et al. \citep{Singh2021} compared the reduced model equation \eqref{eq:reducedmodel} with a problem of perimeter minimisation with constant polarisation patch area, obtaining, for certain initial conditions, a very good agreement between the two solutions.

In the bulk-surface framework, the travelling wave is also influenced by the spatial distribution of the bulk component $b$. In general, given a large difference between $D_b$ and $D_a$, we expect similar dynamics to the reduced model \eqref{eq:reducedmodel}. For instance, on a capsule-shaped domain, when the polarisation patch develops at the center of the cylindrical side, it will slowly move towards one of the two spherical ends \citep{Cusseddu2019}. It remains unclear, however, if and how the bulk component might play a role in pushing the polarised patch towards one of the two spherical ends. 
{\re In the following sections we will compare the two models (BSWP model and its reduced version) on three different geometries. As we will see, in certain cases, the reduced version \eqref{eq:reducedmodel} provides a very good approximation of the overall dynamics of \eqref{eq:bulk}-\eqref{eq:f(a,b)}. However, we will also show that, in other cases, by limiting bulk diffusion, the dynamics can lead to substantially different dynamics.
}

\section{Bulk diffusion and polarisation on an oblate spheroid} \label{sec:curvature-driven polarisation}

\begin{figure}
    \begin{subfigure}[b]{0.4\textwidth}
    \includegraphics[width=1.2\textwidth]{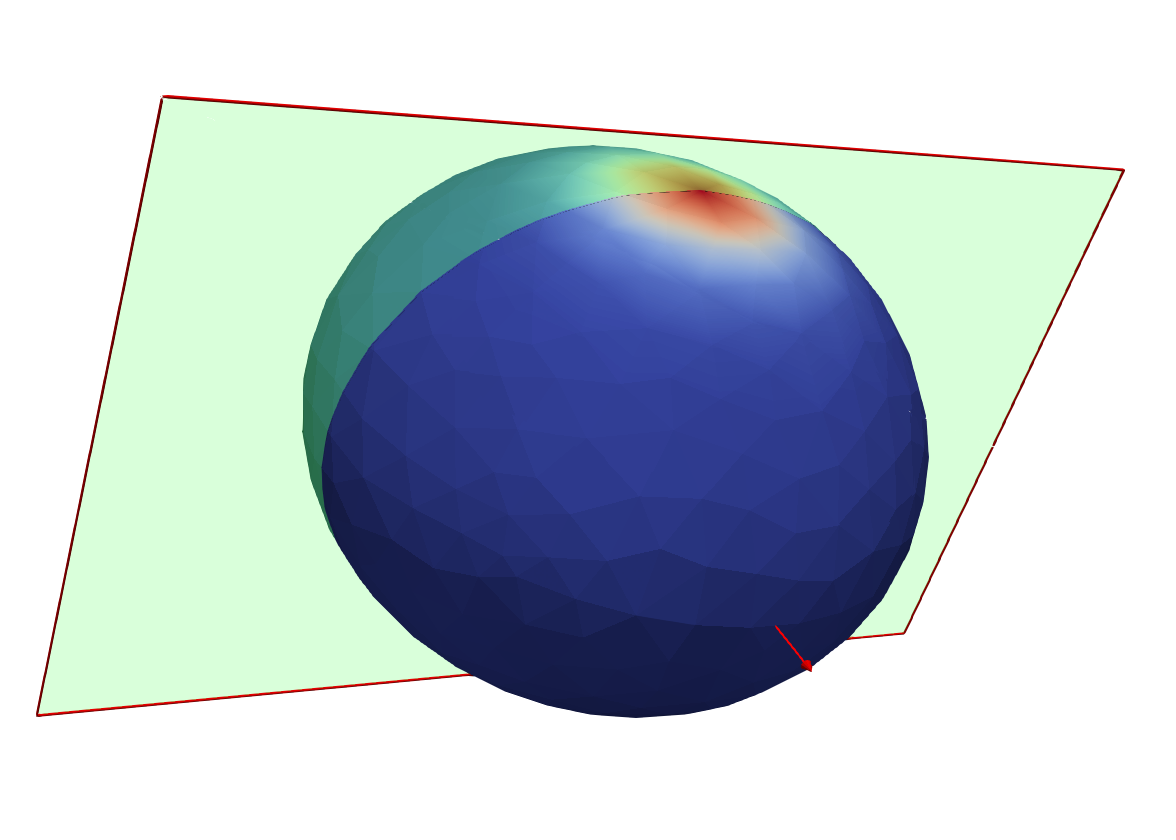}
    \caption{}\label{fig:arclenghtsolutionA}
    \end{subfigure}
    \hfill
    \begin{subfigure}[b]{0.5\textwidth}
    \includegraphics[width=\textwidth]{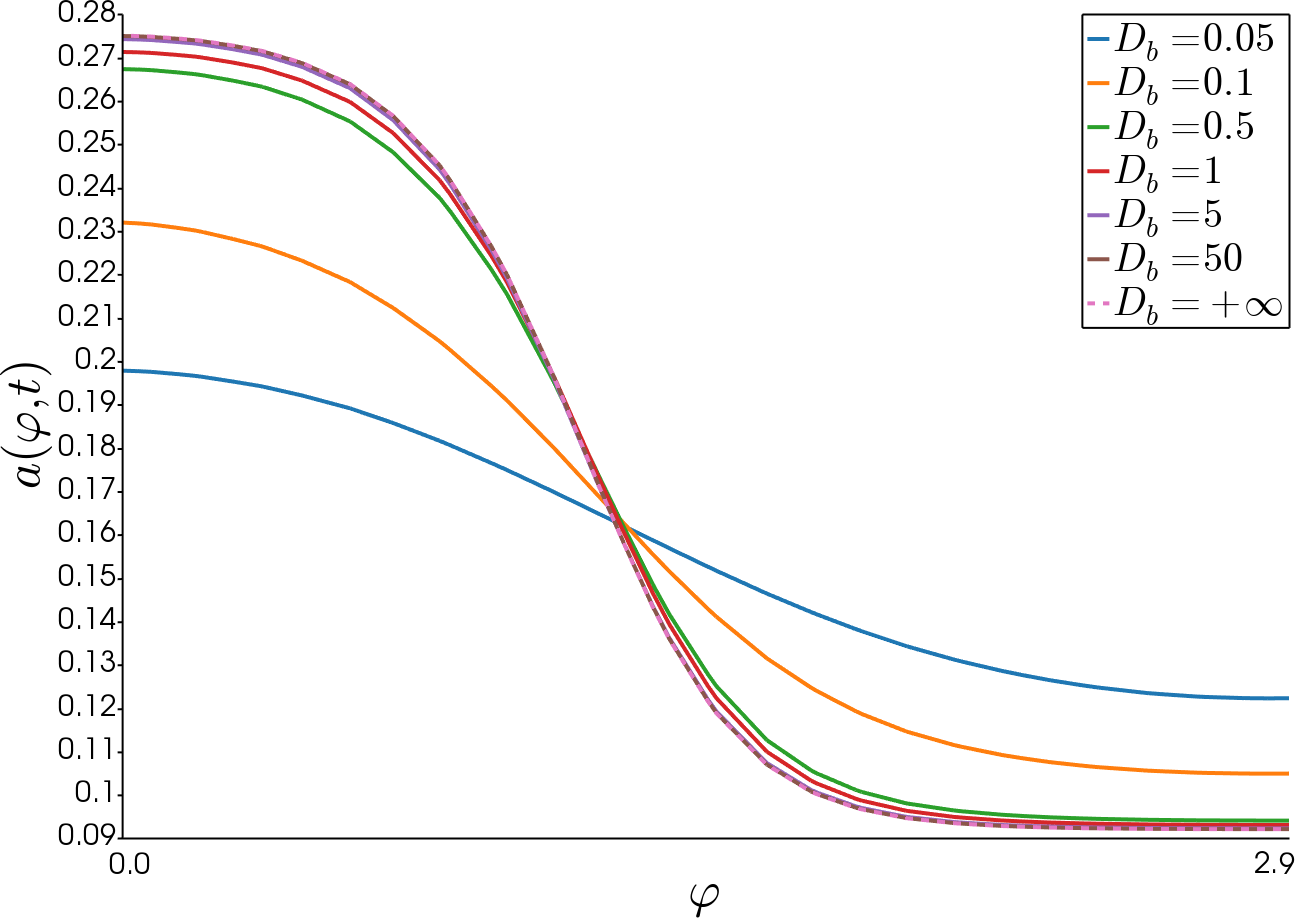}
    \caption{}\label{fig:arclenghtsolutionB}
    \end{subfigure}
    \caption{
    The system \eqref{eq:bulk}-\eqref{eq:f(a,b)} with \eqref{eq:ic_a}-\eqref{eq:ic_b} is solved on a oblate spheroid, for different values of the bulk diffusion coefficient $D_b$. 
    Thanks to the symmetrical properties of $\Omega$
    {\re and of the initial condition for $a(\mathbf{x},t)$ given in \eqref{eq:ic_a}, {\re at the} initial stage, the peak of $a$ expands from the top, symmetrically with respect to all longitudes, over the surface $\Gamma$. This allows us to} plot the solution profiles, {\re for different values $D_b$}, 
    {\re over a meridian,}
    see panel (a). In (b) the profiles are compared at time $t=550${\re, where the variable $\varphi$ represents the meridian length. Therefore $\varphi=0$
    indicates the north pole, $\varphi\approx 2.9$ the opposite pole}. For $D_b=5$ and $D_b=50$ the profile of $a(\mathbf{x},t)$ is almost indistinguishable from the one obtained by the reduced model \eqref{eq:reducedmodel}  (dashed line). It is sufficient to set $D_b = 0.5$ (100 times bigger than $D_a = 0.005$) to obtain a very similar profile as the one from the reduced model. Polarisation occurs also when $D_b = 0.1 = 20 D_a$. For a three-dimensional view of the solutions, see Figure \ref{fig:3D-different-D_bulk}. Parameter values are reported in \ref{app:numerical}.}
    \label{fig:arclenghtsolution}
\end{figure}

In this section, we compare the solution of the reduced model \eqref{eq:reducedmodel} with the solution of the bulk-surface system \eqref{eq:bulk}-\eqref{eq:f(a,b)} for different values of its diffusion coefficient $D_b$ on a simple domain $\Omega\subset\mathbb{R}^3$. 
In order to reduce the complexity, we keep the geometry of $\Omega$ as simple as possible. In the simulations shown here, $\Omega$ is {\re an ellipsoid,}
obtained by a 85\% rescaling of the unit ball $B_1(\mathbf{0})$ on the $y$-axis. In this way, the curvature of $\Gamma$ is maximal for $y=0$ and minimal at {\re at the top and bottom points $(0,\pm0.85,0)$}. {\re When the Gaussian function \eqref{eq:ic_a} is centered at any of these two points, our  choice of the geometric domain} is convenient {\re for }  comparing different solutions of $a$ over a line, obtained by the intersection of $\Gamma$ with a plane passing orthogonal through the peak of the initial condition \eqref{eq:ic_a}{\re, see Figure \ref{fig:arclenghtsolutionA}}. 

\begin{figure}
    \includegraphics[width=1\textwidth]{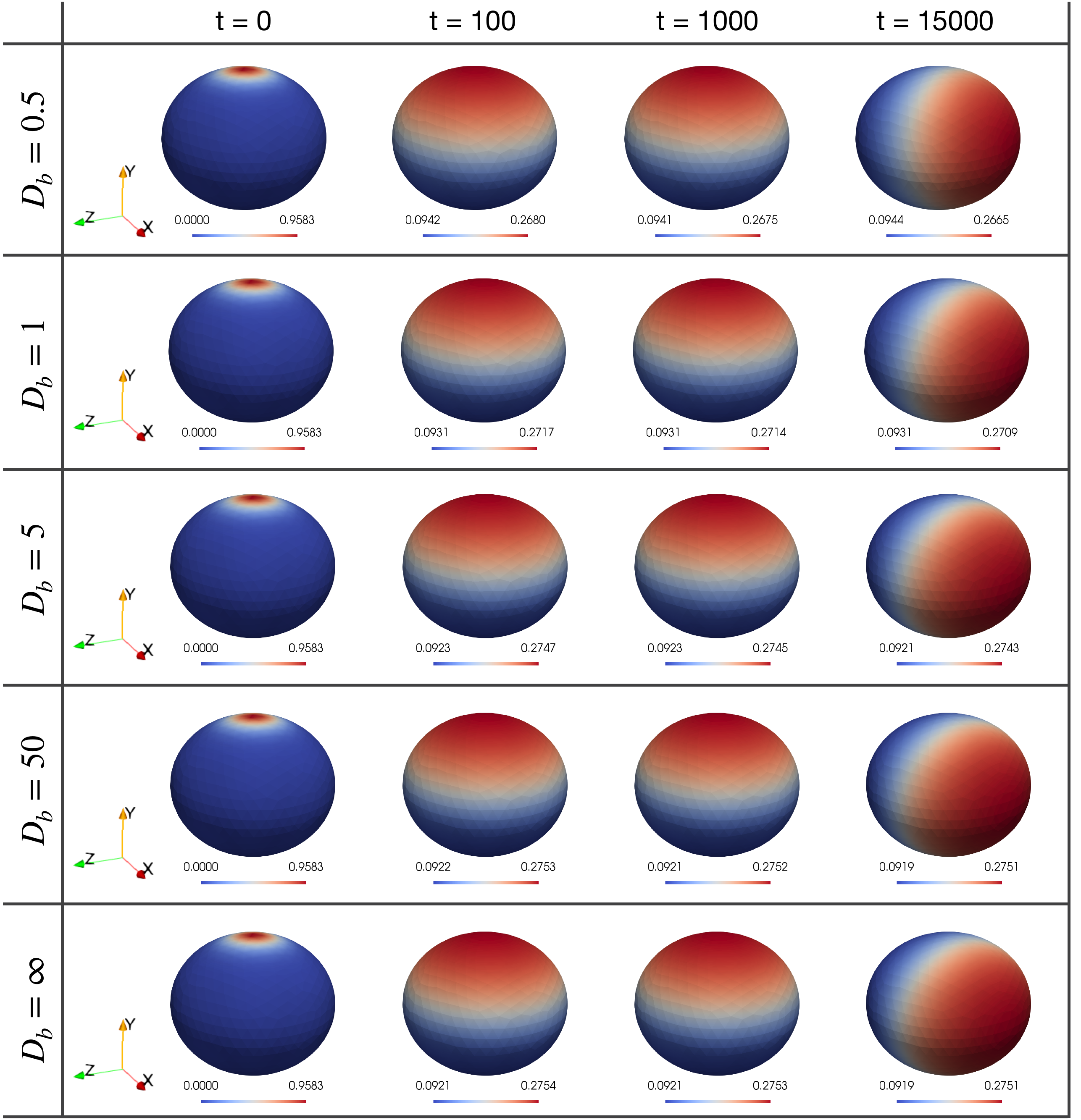}
    \caption{Each row reports the solution $a(\mathbf{x},t)$ of \eqref{eq:bulk}-\eqref{eq:f(a,b)} with \eqref{eq:ic_a}-\eqref{eq:ic_b} for the specified value of the bulk diffusion coefficient $D_b$, at four different times. When $D_b = +\infty$, $a(\mathbf{x},t)$ solves \eqref{eq:reducedmodel} with \eqref{eq:ic_a}. Initial conditions and view point, as shown by the coordinate system on the left, are the same ones for all the plots. The colour bar is automatically re-scaled between $\min_\Gamma a(\mathbf{x},t)$ and $\max_\Gamma a(\mathbf{x},t)$ for each plot.   Parameter values are reported in \ref{app:numerical}. 
    See also Figure \ref{fig:arclenghtsolution}.
    }
    \label{fig:3D-different-D_bulk}
\end{figure}

Figures \ref{fig:arclenghtsolution} and \ref{fig:3D-different-D_bulk} show the solution $a$ corresponding to the bulk-surface wave pinning model \eqref{eq:bulk}-\eqref{eq:f(a,b)}  for different values of $D_b$, and comparisons with the solution of the reduced surface reaction-diffusion model \eqref{eq:reducedmodel} ($D_b=\infty$).
Figure \ref{fig:arclenghtsolutionB} represents the comparison at an intermediate time, when the polarisation patch is still centered at $(x_0,y_0,z_0)$. {\re Consistently with the work by Miller et al. \citep{Miller2022}, the polarisation patch shifts over $\Gamma$ towards the equator, where it stabilises. This behaviour is consistent across different bulk diffusion coefficients, as illustrated in Figure \ref{fig:3D-different-D_bulk}.  As such, the reduced model \eqref{eq:reducedmodel} is a very good approximation of the BSWP model \eqref{eq:bulk}-\eqref{eq:f(a,b)}, for a big variety of bulk diffusion coefficients $D_b$.} 
Increasing $D_b$ causes an enlargement of the distance between maximal and minimal value of $a$, by increasing the first one and decreasing the latter one (Figure \ref{fig:arclenghtsolutionB}). Intuitively, this result can be understood in light of the stability theory developed by Brauns et al. for reaction-diffusion systems of two components with conservation of total mass. We refer the interested reader to their work \citep{Brauns2020} for further details. 

\begin{figure}
    \begin{subfigure}[b]{0.475\textwidth}
    \includegraphics[width=1.\textwidth]{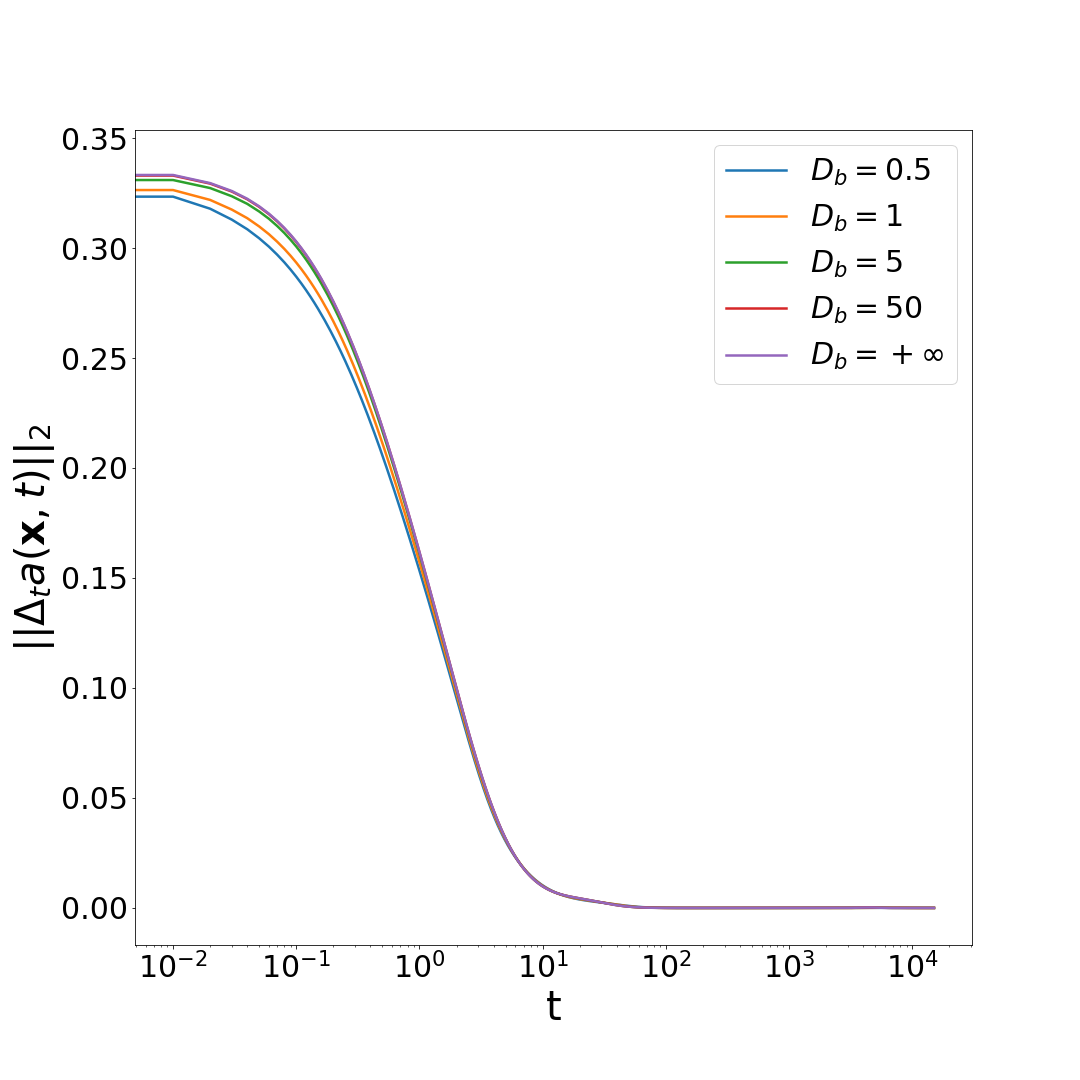}
    \caption{}\label{fig:L2norm-oblate-bulk-diffusion-A}
    \end{subfigure}
    \hfill
    \begin{subfigure}[b]{0.475\textwidth}
    \includegraphics[width=1.1\textwidth]{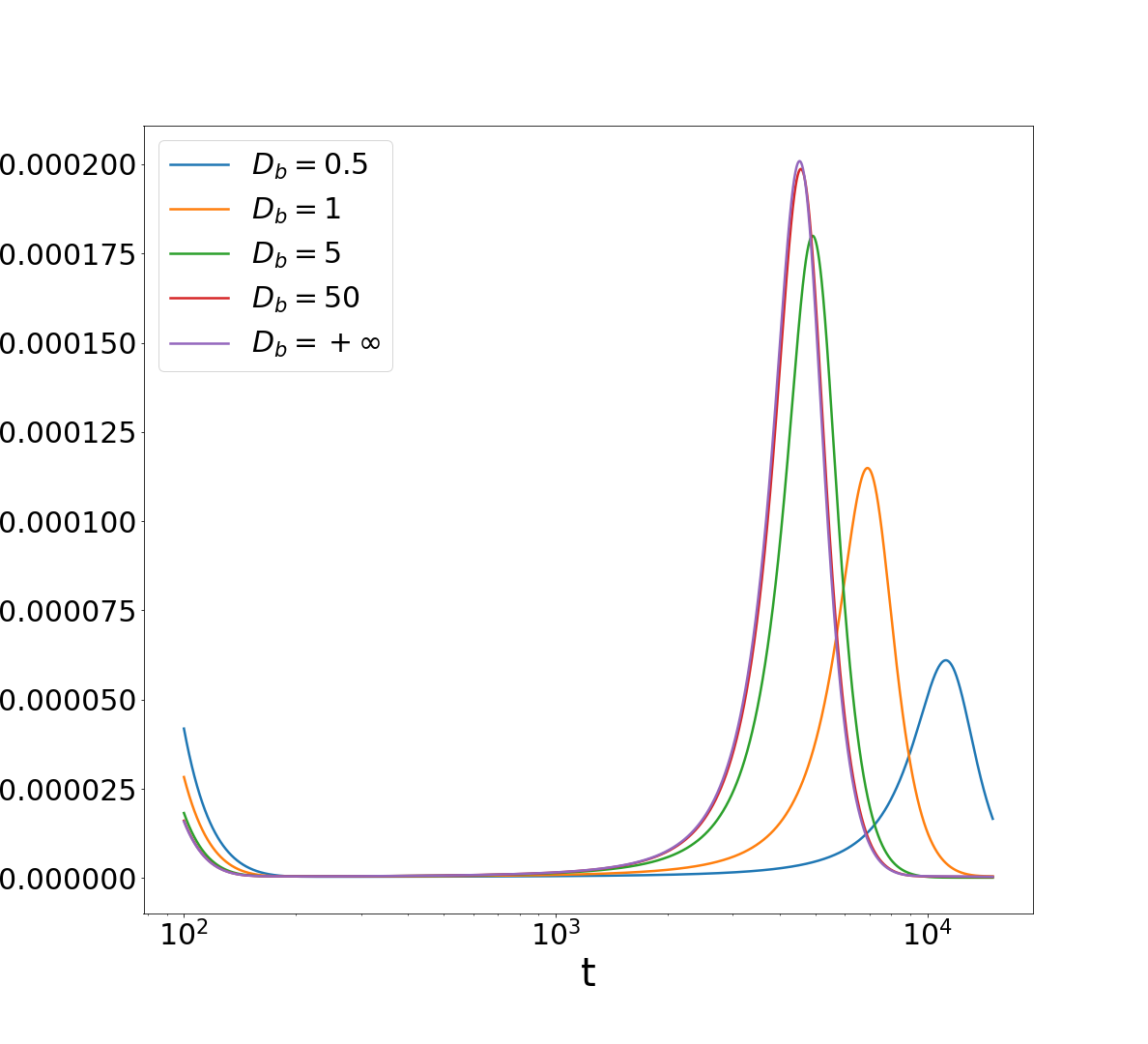}
    \caption{}\label{fig:L2norm-oblate-bulk-diffusion-B}
    \end{subfigure}
    \caption{\re
    Plots of the discrete L$^2$-norm of the function $\Delta_t a(\mathbf{x},t):= a(\mathbf{x},t+\tau) - a(\mathbf{x},t)$, where $a(\mathbf{x},t)$ solves the BSWP model \eqref{eq:bulk}-\eqref{eq:f(a,b)} with \eqref{eq:ic_a}-\eqref{eq:ic_b}, for different values of the bulk diffusion coefficient $D_b$. When $D_b = + \infty$, $a(\mathbf{x},t)$ solves the reduced model \eqref{eq:reducedmodel} with  \eqref{eq:ic_a}. The function $\Delta_t a(\mathbf{x},t)$ represents the difference between consecutive numerical solutions, whose L$^2$-norm noticeably drops in time as the system approaches the polarisation pattern. Panel (b) is a the restriction of the plot in (a) to $t>100$, which highlights an increment of $||\Delta_t a(\mathbf{x},t)||_2$. Such increment is associated with the transition of the polarised pattern across the surface $\Gamma$, see also Figure \ref{fig:3D-different-D_bulk}. The bell profiles in panel (b) would be unnoticeable in scale of panel (a), as their values are very small. }
    \label{fig:L2norm-oblate-bulk-diffusion}
\end{figure}

{\re While the dynamics are very similar for the different values of the bulk diffusion coefficient $D_b$, by varying this parameter we affect the shifting time of the surface polarisation patch towards the equator.} In  
Figure \ref{fig:L2norm-oblate-bulk-diffusion}, the discrete L$^2-$norm of the difference between consecutive numerical solutions $a_h$, defined by 
\begin{equation}
    \label{eq:Delta_t(a)}
||\Delta_t a(\mathbf{x},t) ||_2^2 := \int_\Gamma \left(a_h(\mathbf{x},t+\tau) - a_h(\mathbf{x},t)\right)^2\dsigma,
\end{equation}
where $\tau$ is the temporal discretisation step,  helps in analysing the stability of the system. In particular, it helps to understand how the surface component $a$ is subject to the curvature-driven motion  across the domain $\Gamma$. 
Initially the polarisation patch forms and gets pinned. This causes a drop in $||\Delta_t a ||_2$ and
the  system appears to have reached a steady state (Figure \ref{fig:L2norm-oblate-bulk-diffusion-A}). 
However, after a certain time, $||\Delta_t a ||_2$ starts growing again (Figure \ref{fig:L2norm-oblate-bulk-diffusion-B}), as the polarisation patch shifts from the top to a lateral side of $\Gamma$ (see last column in Figure \ref{fig:3D-different-D_bulk}). The bell shape of $||\Delta_t a ||_2$ (Figure \ref{fig:L2norm-oblate-bulk-diffusion-B})
 indicates that the speed of such transition is not monotonic, and the decreasing side of the profile of $||\Delta_t a ||_2$ indicates system stabilisation.
The bell amplitude indicates the temporal extent of such transition, which occurs on a much longer time scale with respect to the initial propagation and pinning of the polarised patch. It must be noted that time in Figure \ref{fig:L2norm-oblate-bulk-diffusion} is reported on a log scale.
It is also interesting to remark that the profiles in Figure \ref{fig:L2norm-oblate-bulk-diffusion-B} are not visible in Figure \ref{fig:L2norm-oblate-bulk-diffusion-A}, because the maximal speed of such transition is extremely small, when compared to the initial propagation (see also $y$-axis values in both figures). 

{\re In the case here presented, the bulk diffusion coefficient $D_b$ seems to have a role only on the temporal scale, meaning that, when  bulk diffusion is limited, the surface patch is subject to a slower curvature-dependent speed. However, we will now present another case in which the parameter $D_b$ appears to have a much stronger impact on the model dynamics.
}

\section{Competition between different polarisation patches}\label{sec:competition}

In \citep{Cusseddu2019} we showed an example in which an initial condition for $a(\mathbf{x},t)$ with two separated peaks, subject to the BSWP model \eqref{eq:bulk}-\eqref{eq:f(a,b)}, evolved, in a first phase, into two polarisation patches. However, in the long time, eventually one of the two prevailed over the other.
Competition between multiple polarisation patches in the solution $a(\mathbf{x},t)$ has been studied by Chiou et al. \cite{Chiou2018} also for the limit case $D_b=+\infty$ (i.e. equation \eqref{eq:reducedmodel}). They show how different patches compete on a flat surface, where, as discussed in Section \ref{sec:model_reduction}, the speed of the travelling wave is given by $v = c - \kappa D_a$, with $\kappa$ being the curvature of the patch interface. Therefore, in the case of two circular polarisation patches of radii, respectively, $R$ and $r$ with $R > r > \frac{D_a}{c(b_0)}$,
they will both spread radially as long as
$v_R = c(b) - \frac{D_a}{R}$ and $v_r = c(b) - \frac{D_a}{r}$ remain positive. However,
while $a(\mathbf{x},t)$ propagates, $b(\mathbf{x},t)$ gets depleted and $c$ decreases. This continues until a certain point is reached such that $c(b) < \frac{D_a}{r}$. At this point, $v_R$ remains positive, but for the smaller patch the speed is reversed and it starts shrinking. The shrinking patch releases $b(\mathbf{x},t)$, which, in the case of $D_b=+\infty$, is immediately consumed by the other enlarging patch \citep{Chiou2018}. However, in the bulk-surface framework, the component $b(\mathbf{x},t)$ is released locally, not globally.  
Therefore, the bulk-surface wave pinning model might develop different competition dynamics from those of the reduced model \eqref{eq:reducedmodel}. 
In the case of $\Gamma\in\mathbb{R}^1$, the situation is different, as such speed description does not hold. However, peak competition is not specific to multidimensional domains, as shown and analysed on one-dimensional domains in \citep{Chiou2018, Brauns2021-wavelength}.

\begin{figure}
\includegraphics[width=1\textwidth]{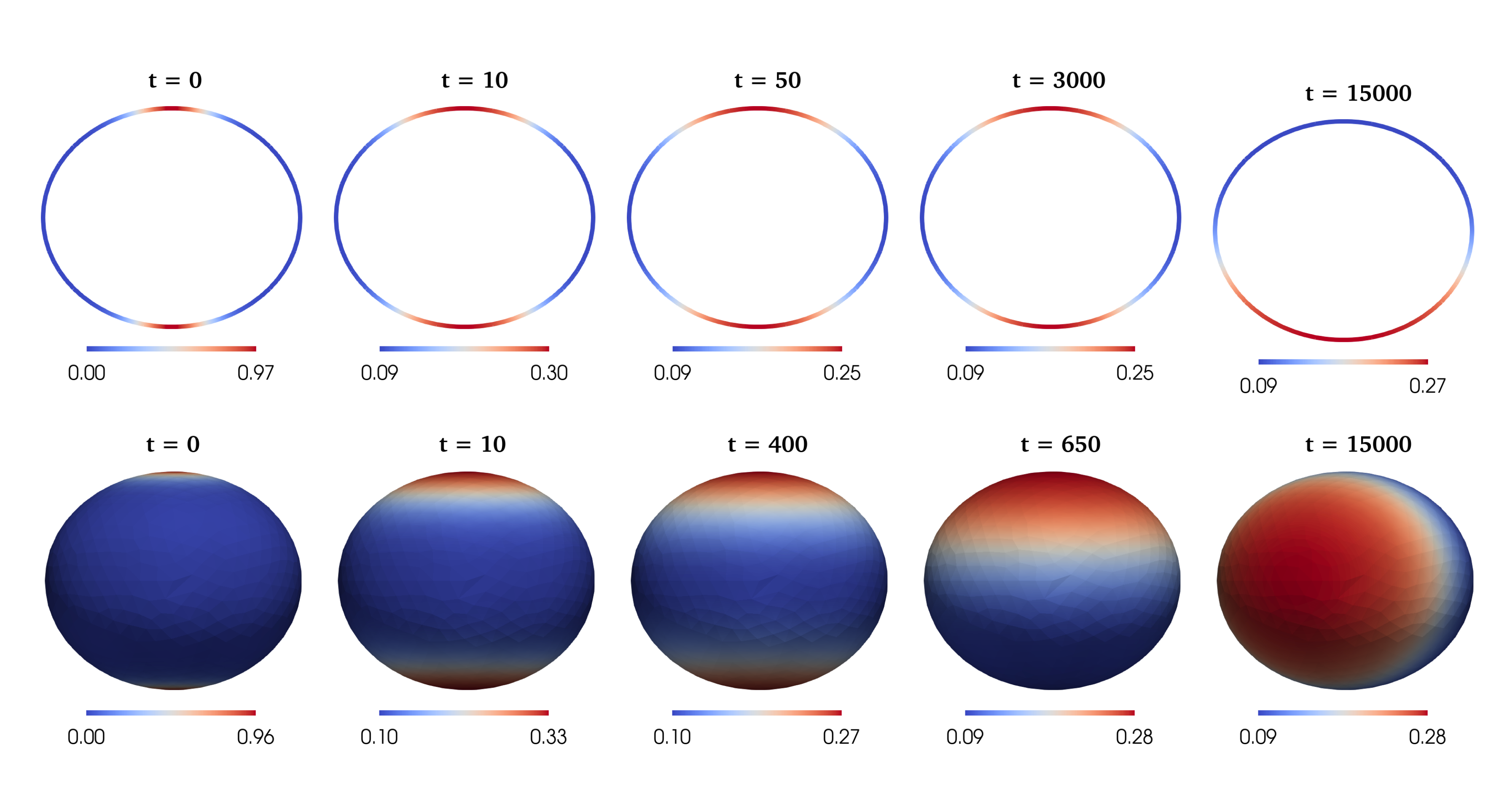}
    \caption{The solution $a(\mathbf{x},t)$ corresponding to the reduced surface reaction-diffusion model \eqref{eq:reducedmodel} on a 
    curve and surface, with initial condition  \eqref{eq:TBC_ic}, where $x_0=x_1=z_0=z_1=0$ and $y_0=-y_1 = 0.85$. The colour bar is automatically re-scaled between the lowest and highest value of the solution at each plot. In the top row $M_0=0.85|\Omega|$, while on bottom row $M_0=|\Omega|$. Remaining parameter values are reported in \ref{app:numerical}.
    }
    \label{fig:TBC-3D_and_2D}
\end{figure}
\begin{figure}
\includegraphics[width = 1\textwidth]{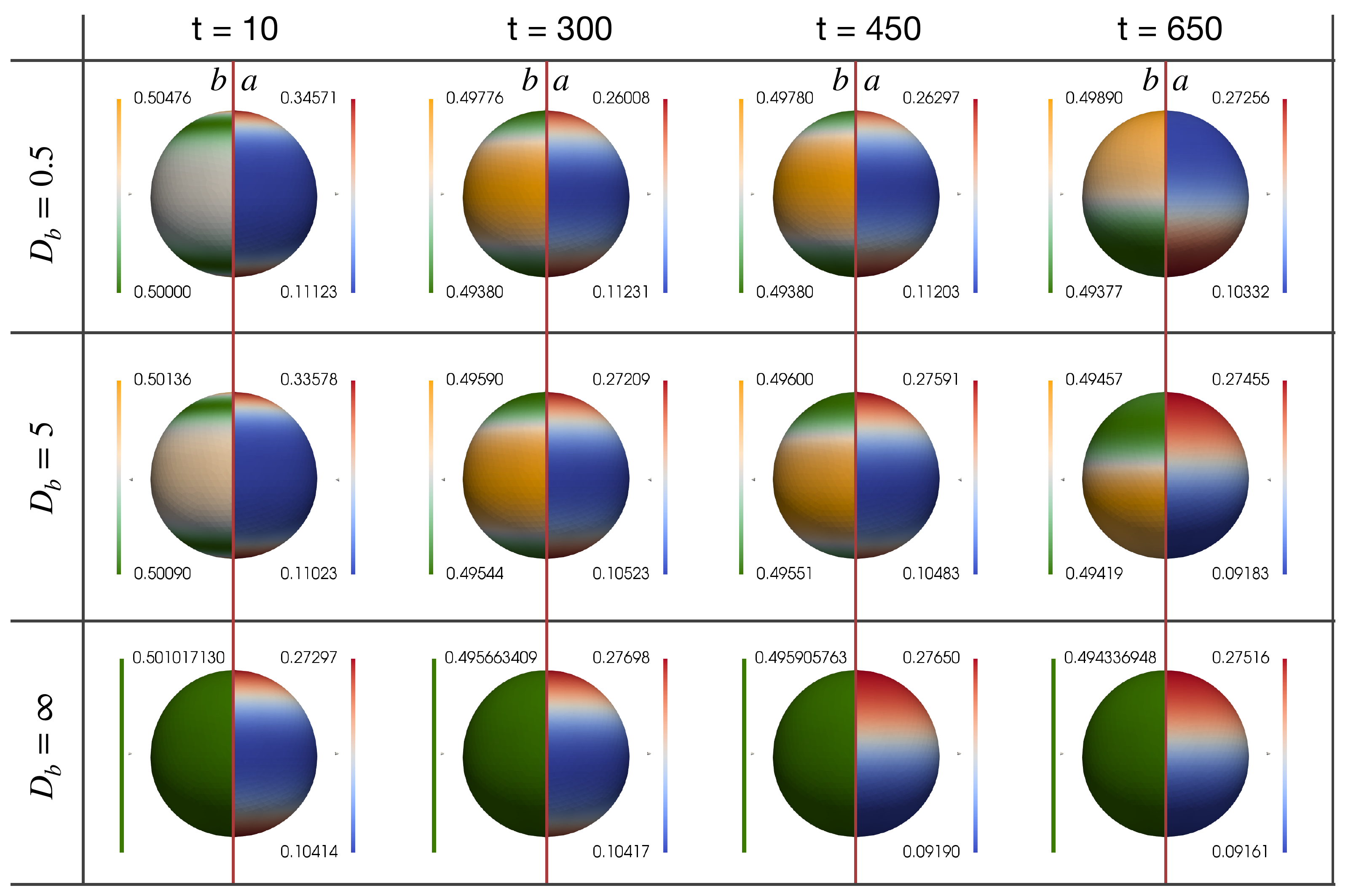}
    \caption{
    The solutions $a(\mathbf{x},t)$ and $b(\mathbf{x},t)$ of the BSWP model \eqref{eq:bulk}-\eqref{eq:f(a,b)} with initial conditions \eqref{eq:ic_b} and \eqref{eq:TBC_ic} 
    {\re for $D_b = 0.5$ and $D_b = 5$ (first and second row)
    and the solution of \eqref{eq:reducedmodel} 
     with initial condition \eqref{eq:TBC_ic} (indicated by $D_b = +\infty$, last row). Columns indicate the solutions at different times, i.e. $t=$ 10, 300, 450, 650.
     We show both bulk and surface solution by longitudinally sectioning the domain in two halves: the bulk component $b$ is reported on the left half (color map green-orange), the surface component $a$ on the right half (color map blue-red). For simplicity of comparisons, the case $D_b = \infty$ is represented as the previous ones, but $b$ is exactly $(M_0 - \int_\Gamma a)/|\Omega|$. 
     While for $D_b = 0.5$ the bottom patch wins the competition, when increasing such value, the outcome is the opposite. These results refer to the mesh shown Figure \ref{fig:meshes}F. Here $D_a = 0.005$, while the remaining parameter values are reported in  \ref{app:numerical}. A video of the simulations presented here is attached as supplementary material, together with the same simulations for less refined mesh.}
    }
    \label{fig:TBC}
\end{figure}

In Figure \ref{fig:TBC-3D_and_2D}, we show the solution of the surface reaction-diffusion \eqref{eq:reducedmodel} (the reduced model), given the initial condition
\begin{equation}\label{eq:TBC_ic}
    a(\mathbf{x},0) = \sum_{i=0}^1 a_{p,i}\exp{\left\{-{\sigma_{x,i}^{-2}}{(x-x_i)^2} - {\sigma_{y,i}^{-2}}{(y-y_i)^2} - {\sigma_{z,i}^{-2}}{(z-z_i)^2}\right\}},
\end{equation}
where $(x_0,y_0,z_0)$ and $(x_1,y_1,z_1)$ are, respectively, the points with the smallest curvature, and we would refer to these as the 
top and bottom of $\Gamma$. When $\Gamma$ is a surface (bottom row of Figure \ref{fig:TBC-3D_and_2D}), the patch competition is immediately followed by the movement of the winner towards the central area of the domain. When $\Gamma$ is a curve (top row of Figure \ref{fig:TBC-3D_and_2D}), the solution stabilises immediately after patch competition. In this last case, the curve $\Gamma$ {\re is an ellipse} obtained by a 85\% re-scaling on the $y$-axis of the circumference of radius one and can be seen as the intersection of the surface on the bottom row of the figure with the plane $z=0$. Hence, the initial condition for $a(\mathbf{x},t)$ is given by \eqref{eq:TBC_ic} with $z=z_0=z_1=0$. For numerical reasons due to the {\re three-dimensional} meshing, despite the initial condition being symmetric along the $y$-axis, the mass of the peak centered at the top $(x_0,y_0,z_0)$ results to be slightly bigger than the mass of the other peak (approximately 0.30498 versus 0.30476). This creates a natural perturbation that triggers the competition.  Brauns et al., in \citep{Brauns2021-wavelength}, provide arguments to show the ultimate instability of multiple peak solutions to two-component mass conserving reaction-diffusion systems. In particular, they show how the difference of mass between two peaks will always grow, leading to the disappearance of one patch. {\re  
Hence, following \citep{Brauns2021-wavelength}, we might expect the peak at the top to expand at the expenses of the other one. {\re  The bottom row in Figure \ref{fig:TBC-3D_and_2D} shows that this is indeed the case. Would this be true also for the complete BSWP system \eqref{eq:bulk}-\eqref{eq:f(a,b)}? } 

In order to address this question and to investigate the role that bulk diffusion plays in the competition between multiple polarisation patches, we consider another relatively simple domain $\Omega\subset\mathbb{R}^3$. Here $\Omega$ results from a small ``egg-type'' deformation of the unit sphere, obtained by a domain rescaling along the $y$-axis: the hemisphere $\Omega\cap \{y>0\}$ is stretched by $5\%$, while the hemisphere $\Omega\cap \{y<0\}$ is shrunk by $5\%$. As such, the point $(0, 1.05, 0)$ is the point with maximal curvature over $\Gamma$, while $(0, -0.95, 0)$ the one with minimal curvature. We might refer to these points as the
top and bottom of $\Gamma$. 
In Figure \ref{fig:TBC} we show the solutions $a$ and $b$ of the BSWP model \eqref{eq:bulk}-\eqref{eq:f(a,b)} with initial conditions \eqref{eq:ic_b} and \eqref{eq:TBC_ic}. Again, the initial condition for $a$ is given by the sum of two Gaussian functions centered, respectively, at the top and bottom of $\Gamma$. In these simulations, the peak at the flattened bottom hemisphere has a slightly bigger mass than the top peak. However, in this case, competition might not depend only on this property. Hence, as already discussed, the expansion of the initial peaks strictly depends also on the curvature of the domain. Indeed, following the work by Singh et al. \cite{Singh2021} we might expect the model to evolve towards a single patch solution that minimises its interface perimeter. Interestingly, the bulk diffusion coefficient $D_b$ reveals to be crucial in determining the outcome of the competition. In the reduced model \eqref{eq:reducedmodel} the bottom patch looses the competition and the polarisation occurs at the top, where the curvature is greater. The same result is achieved by the complete system \eqref{eq:bulk}-\eqref{eq:f(a,b)} when the bulk diffusion coefficient is big enough. It might be worth observing that, in these two similar cases, bulk diffusion slows down the competition. 
However, by further reducing the bulk diffusivity, the BSWP model \eqref{eq:bulk}-\eqref{eq:f(a,b)} results in the opposite outcome, with  polarisation of the bottom hemisphere, at the expense of the top one. This result is confirmed on three different mesh refinements, those results are attached as supplementary material. 

We have here shown the first crucial difference that might arise in the BSWP model when $D_b$ is not extremely big. It is important to stress that, by considering $D_b=0.5$, we are still considering it to be 100 times bigger than the surface diffusion coefficient, which is $D_a = 0.005$. The remaining parameter values are reported in \ref{app:numerical}.
}

\section{Bulk heterogeneity can sustain cell polarisation}\label{sec:bulk induces polarisation}
\begin{figure}
    \includegraphics[width=1.0\textwidth]{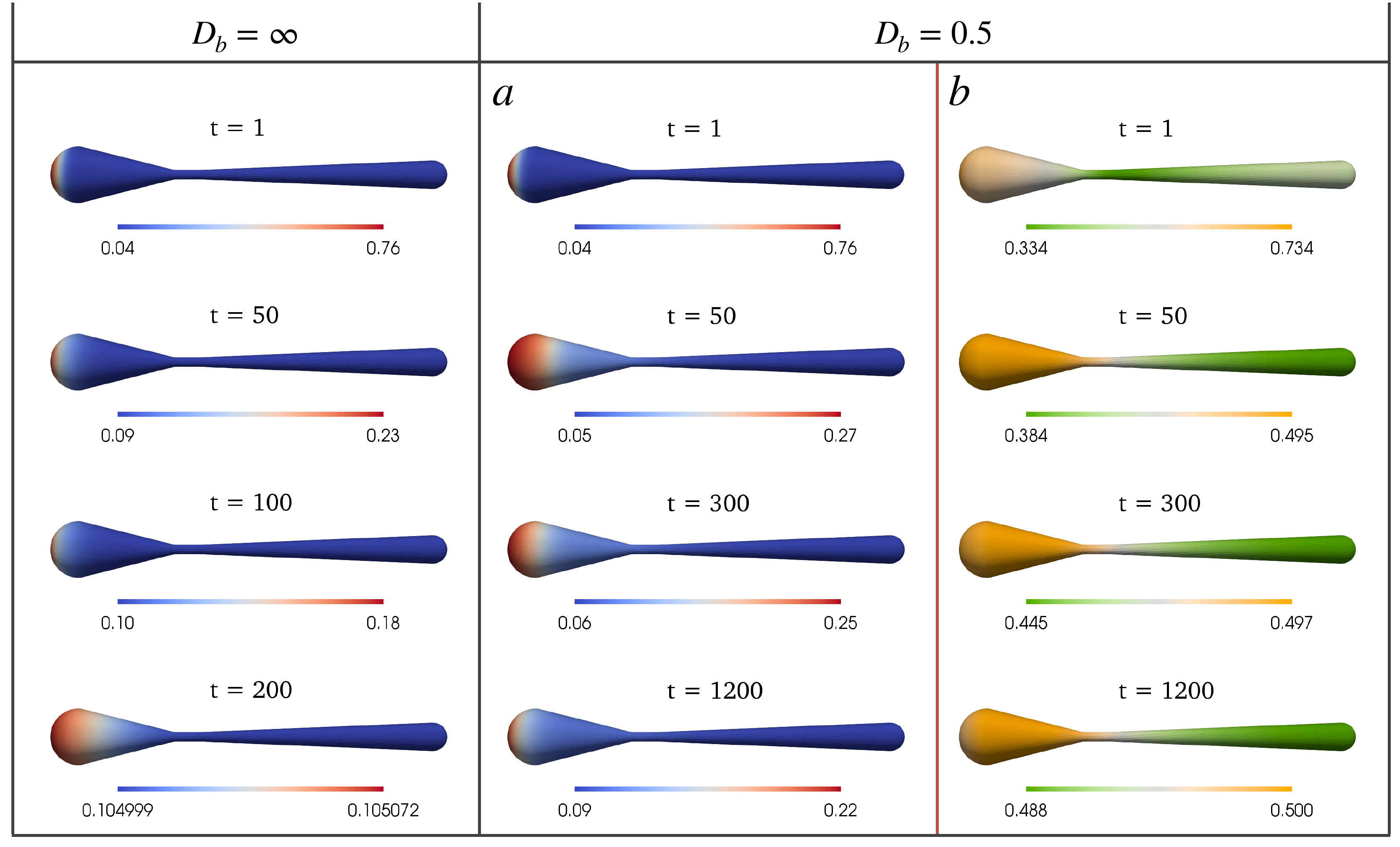}
\caption{Bulk polarisation induces surface polarisation: here the reduced surface reaction-diffusion model \eqref{eq:reducedmodel}-\eqref{eq:ic_a} does not generate a lasting polarisation (first column), as compared to the BSWP model \eqref{eq:bulk}-\eqref{eq:f(a,b)} with \eqref{eq:ic_a}-\eqref{eq:ic_b}. {\re Notice that, while the solution of the reduced model at time $t=200$ might appear to generate polarisation, the difference between minimum and maximum is extremely minimal.} Bulk-surface finite element solutions $a(\mathbf{x},t)$ and $b(\mathbf{x},t)$ are reported, respectively, in the second and third column. The domain elongates along the $y$ axis and the initial condition \eqref{eq:ic_a} for $a(\mathbf{x},t)$ is centered at the smallest $y$ value. In both cases we set $M_0=1.2|\Omega|$. The BSWP model is solved for $D_b=0.5$.  For the remaining parameter values see \ref{app:numerical}.}
    \label{fig:nonconvex}
\end{figure}

\begin{figure}
\includegraphics[width=1.0\textwidth]{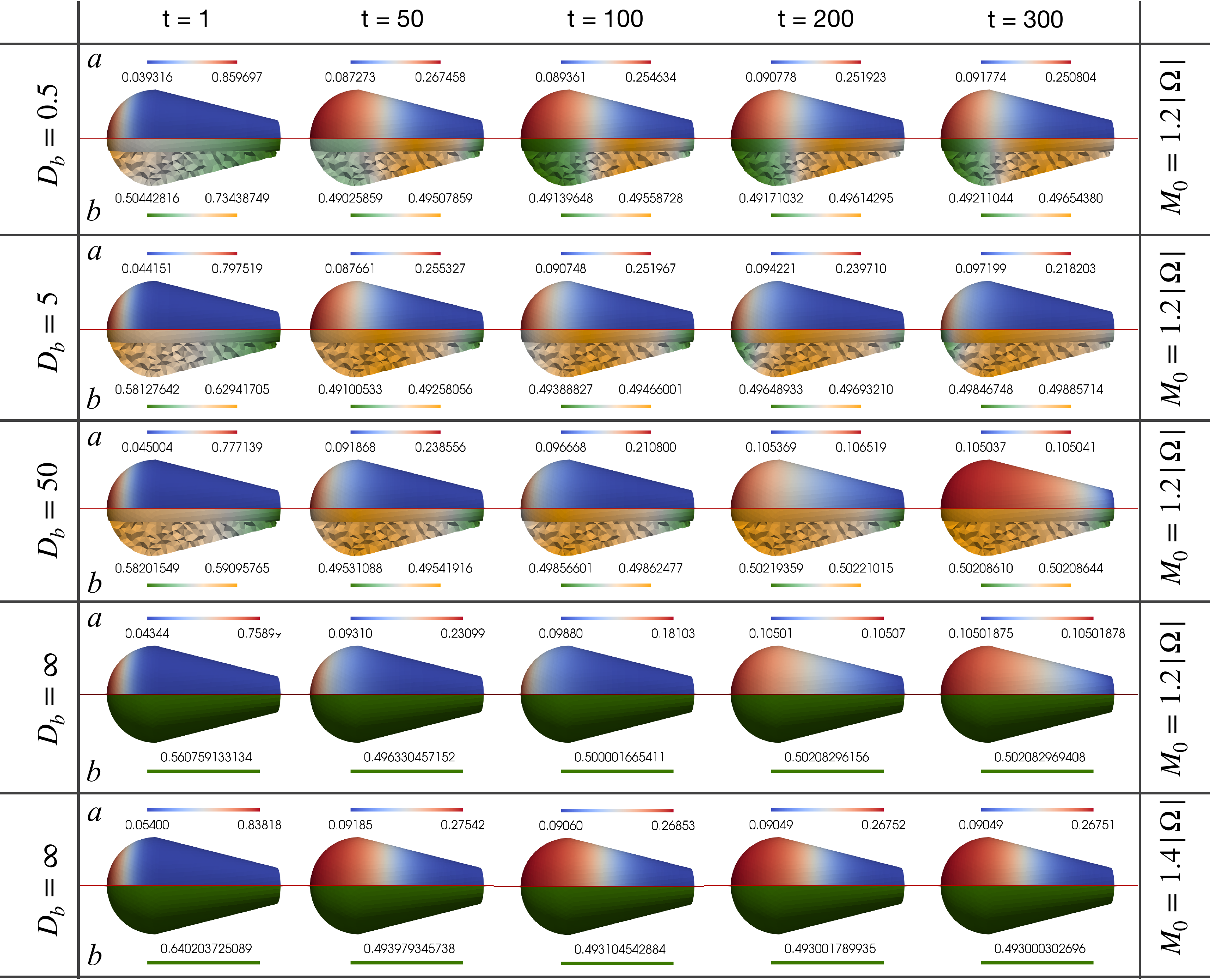}
\caption{\re 
A zoom on the simulations of Figure \ref{fig:nonconvex} helps in understanding the polarisation dynamics. Here, we show the numerical solutions on a portion of the domain $|\Omega|$ where the surface component has its initial peak. The first three rows of figures represent the solutions of the BSWP system \eqref{eq:bulk}-\eqref{eq:f(a,b)} with \eqref{eq:ic_a}-\eqref{eq:ic_b} for, respectively, bulk diffusion $D_b = 0.5$, 5 and 50. The last two rows represent the solutions of the reduced model \eqref{eq:reducedmodel}-\eqref{eq:ic_a}
with different total mass, the last one having total mass $M_0 = 1.4|\Omega|$, instead of $M_0 = 1.2|\Omega|$. Each single figure shows both the surface and bulk solutions, separated by an horizontal red line cutting the domain section: the surface component $a$ is shown above the line (blue-red colourmap), while below the line we report the bulk component $b$, with a view also on its interior (green-orange colourmap). Each colourbar is automatically rescaled between the minimum and maximal solution value achieved in such section of $\Omega$. 
Surface polarisation is strictly dependent on the bulk component in the vicinity of the polarisation interface. By increasing the total mass from $M_0 =1.2|\Omega|$ to $M_0 =1.4|\Omega|$, polarisation results also in the reduced model. When $M_0 =1.2|\Omega|$, the BSWP model with a sufficiently small $D_b$ induces spatial dishomogeneities in the bulk component that are able to locally sustain surface polarisation. 
}\label{fig:zoom_on_cones}
\end{figure}

Until now, our discussion has been focused on numerical investigations of the bulk-surface wave pinning model on a relatively simple convex domain. In this section we consider a more complex three-dimensional geometry, taking advantage of the bulk-surface finite element method which is easily adaptable to different geometries (see \ref{app:numerical} and \citep{Cusseddu2019, CussedduPhD, MADZVAMUSE20169} for details). 
In a bulk-surface framework, local restrictions of the domain $\Omega$ slow down bulk diffusion, maintaining heterogeneity in the bulk component for a longer time. This might be fundamental for a lasting polarised pattern. In a non-convex geometry we find that the impact of a slower bulk diffusion might be crucial to polarisation. In Figure \ref{fig:nonconvex}, we show one example of transient polarisation. Here, neither the reduced surface reaction-diffusion model \eqref{eq:reducedmodel} nor the BSWP model \eqref{eq:bulk}-\eqref{eq:f(a,b)} are able to maintain a stable and strong polarisation of the surface $\Gamma$. However, in 
both cases, {\re we notice the presence of a polarisation patch}, but in
the case of finite bulk diffusion, it resists for a much longer time. 
In the case of the reduced surface reaction-diffusion model \eqref{eq:reducedmodel}, {\re we can say that} polarisation is completely lost at $t=200$, while the BSWP model \eqref{eq:bulk}-\eqref{eq:f(a,b)} still maintains strong polarisation at $t=1200$ {\re (the difference between the extreme values $a_{high}$ and $a_{low}$ reported on each colourbar in the two cases gives an idea on the polarisation level)}.
This is due to the bulk patterning, shown in the third column of Figure \ref{fig:nonconvex}. Both slow diffusion ($D_b=0.5$) and domain geometry highlight the importance that the bulk component $b(\mathbf{x},t)$ has in establishing polarisation patterns. \\
{\re A deeper observation of these dynamics is reported in Figure \ref{fig:zoom_on_cones}. Here we show the solution on a portion of the domain where the initial peak is prescribed. In this way the local values are more easily distinguishable and we can focus on the bulk component $b$ just in the proximities of the polarisation patch interface. Figure \ref{fig:zoom_on_cones} shows the solutions of the BSWP system \eqref{eq:bulk}-\eqref{eq:f(a,b)} for different values of the parameter $D_b$, as well as the solution of the reduced model \eqref{eq:reducedmodel} for two different values of the total mass $M_0$. We show that polarisation is achievable also by the reduced model, at the cost of increasing $M_0$. Therefore, reducing bulk diffusion might be a way of over-passing such cost, since inhibited spatial redistribution of the bulk component might be a way of keeping, in the vicinity of the the polarisation interface, the values of $b$ within a certain range for polarisation.
}

In general, it is worth mentioning that non-convex geometries might be of particular interest in the study of cell polarisation, as they might have a remarkable role in the dynamics. {\re Spill et al. studied a more complex bulk-surface model for cell polarisation and showed the importance of cell geometry in achieving and maintaining polarisation \citep{Spill2016}. In particular, they show how a non-convex cell geometry might be more convenient for polarising, with respect to convex geometries. 
}
Ramirez et al. investigated polarisation of dendritic spines, which are small protrusions present in neuron cells  \citep{ramirez2015dendritic}. In their work they showed how their geometry can, by itself, induce polarisation, even in the absence of depletion of the bulk component. Indeed, propagating fronts of solutions for single bistable reaction-diffusion equations may experience geometry-induced pinning when they reach an abrupt opening of the domain \citep{bialecki2020traveling, ramirez2015dendritic,berestycki2016front}, which is the case of the modelled spines. Moreover,
Giese et al. studied the impact of obstacles in a two-dimensional version of the BSWP model \citep{Giese2015}, observing how these might influence the position of the polarisation patch. From a modelling perspective, this is fundamentally important since in an environment such as that of the cell, crowding effects and sub-cellular structures might finally trigger surprising outcomes.

\section{Conclusion}\label{sec:conclusion}
In this manuscript we presented and compared novel bulk-surface finite element numerical computations concerning the long time dynamics of the bulk-surface wave pinning model \eqref{eq:bulk}-\eqref{eq:f(a,b)} and its natural asymptotic approximation, the surface reaction-diffusion model given in \eqref{eq:reducedmodel}.
While in \citep{Cusseddu2019} we focused on presenting intermediate results on biologically relevant timescales, in this work we consider a rescaled version of the system on simpler domains, with the goal of understanding the role that the bulk component might have on the surface dynamics.

In the bulk-surface system \eqref{eq:bulk}-\eqref{eq:f(a,b)}, the polarisation pattern over the surface $\Gamma$ is strictly dependent on the bulk component $b(\mathbf{x},t)$, especially on its diffusion coefficient $D_b$.
For too small values of $D_b$ polarisation generally cannot occur through wave pinning dynamics. When $D_b$ is large enough, polarisation patterns can be generated and these are subject to a slow transition of the polarised area across the surface $\Gamma$. This is an intrinsic feature of equation \eqref{eq:surface} for $a(\mathbf{x},t)$, as recently shown, for instance by Singh et al. \citep{Singh2021} {\re and Miller et al. \citep{Miller2022}} for the reduced surface reaction-diffusion model \eqref{eq:reducedmodel}.

In our work we discuss the importance of the bulk component $b(\mathbf{x},t)$ in regulating the final position of the polarised patch as well as its transition speed. While the overall system dynamics in the reduced case \eqref{eq:reducedmodel} often constitute a good qualitative representation of the complete system \eqref{eq:bulk}-\eqref{eq:f(a,b)}, considering the role of the bulk component $b(\mathbf{x},t)$ might still be of quantitative importance from the biological point of view of cell polarisation. Moreover, in some cases, transient polarisation can directly result from bulk heterogeneity. 

The reduced surface reaction-diffusion model \eqref{eq:reducedmodel} constitutes an even more minimal model for cell polarisation with respect to the bulk-surface wave-pinning model, however it is important to keep in mind that it is based on the assumption that nothing hinders or limits the bulk diffusion. In some other cases, the assumption that the bulk proteins are so abundant leads to consider the bulk component constant both in space and time. Even with this assumption polarisation can occur, as shown for similar cases in \citep{ bialecki2020traveling,ramirez2015dendritic,berestycki2016front}. 

Clearly, these assumptions depend on the question that a modeller formulates and wants to answer. However, for the case when obstacles exist for the bulk diffusion, such as holes {\re or tethers}, different polarisation dynamics might arise {\re \citep{Giese2015, houk2012membrane}}. In the same way, as we have seen, restrictions of the domain $\Omega$, which might represent contractions of the cell membrane, can induce polarisation through bulk patterning. This suggests that, on a moving and deforming domain, the bulk component might be of fundamental importance.

\section*{Acknowledgments}
This project (DC,AM) has received funding from the European Union's Horizon 2020 research and innovation programme under the Marie  Skłodowska-Curie grant agreement No 642866. DC is supported by Fundação para a Ci\^encia e a Tecnologia under the project 
UIDB/00208/2020. AM acknowledges support from the EPSRC grant (EP/T00410X/1): UK-Africa Postgraduate Advanced Study Institute in Mathematical Sciences (UK-APASI). The work of AM was partly supported by Health Foundation (1902431), the NIHR (NIHR133761) and by an individual grant from the Dr Perry James (Jim) Browne Research Centre on Mathematics and its Applications (University of Sussex). AM acknowledges support from Royal Society Wolfson Research Merit Award (2016-2021) funded generously by the Wolfson Foundation. AM is a Distinguished Visiting Scholar to the Department of Mathematics, University of Johannesburg, South Africa.

\section*{Conflict of interest}
The authors declare no conflict of interest.

\appendix
\section{The bulk-surface finite element method \re and numerical details}\label{app:numerical}
Both systems \eqref{eq:bulk}-\eqref{eq:f(a,b)} and \eqref{eq:reducedmodel} are solved in FEniCS \citep{Alnaes2015} (see code in the supplementary material or online at github.com/davidecusseddu/BWSP22). In the case of the BSWP model \eqref{eq:bulk}-\eqref{eq:f(a,b)} we use the bulk-surface finite element numerical method proposed in \citep{Cusseddu2019}, which combines bulk-surface finite element method in space and semi-implicit finite difference in time. For the spatial discretisation, we generate a mesh $\Tau_h$ over the whole domain $\Omega$, those elements are tetrahedra if $\Omega\subset\mathbb{R}^3$, or triangles if $\Omega\subset\mathbb{R}^2$. A mesh $\Sau_h$ of the surface $\Gamma$ is then naturally induced by the first one, being composed by all the faces or edges of the elements enclosing $\Tau_h$. At each time step we solve three systems of linear equations: firstly a predictor for the solution $a$ is calculated by using an ImEx (implicit diffusion, explicit reactions) scheme. In the successive two steps Crank-Nicolson scheme is used to calculate $b$ and correct $a${\re, which increases the accuracy in time. For more details on the numerical method, we refer the reader to our previous works \cite{Cusseddu2019, CussedduPhD}}. In the reduced case \eqref{eq:reducedmodel}, $b$ is spatially homogeneous, therefore in the second step of the numerical method we calculate its predicted value by applying the conservation of total mass.
{\re Conservation of total mass is a key property of the BSWP model and the numerical scheme that we use keeps such property. 
We show this in Figure \ref{fig:plot_masses}. For all the simulations presented in this manuscript, at each time point we calculate the total mass of the numerical solutions. Among all the simulations the total mass is kept constant up, in the worst cases from Figure \ref{fig:3D-different-D_bulk}, to an order of magnitude of $10^{-9}$. 
}
All the geometries, with the respective meshes as reported in Figure \ref{fig:meshes}, were created using Gmsh  \citep{Geuzaine2009}. Meshes are included in the supplementary material {\re as .xml file, together with a .geo file of the elongated geometry in Figure \ref{fig:meshes} C. 
In Table \ref{tab:discretisation} we report numerical details on the spatial and temporal discretisation of all the cases considered in this paper.}

\begin{figure}
    \centering
    \includegraphics[width=0.9\textwidth]{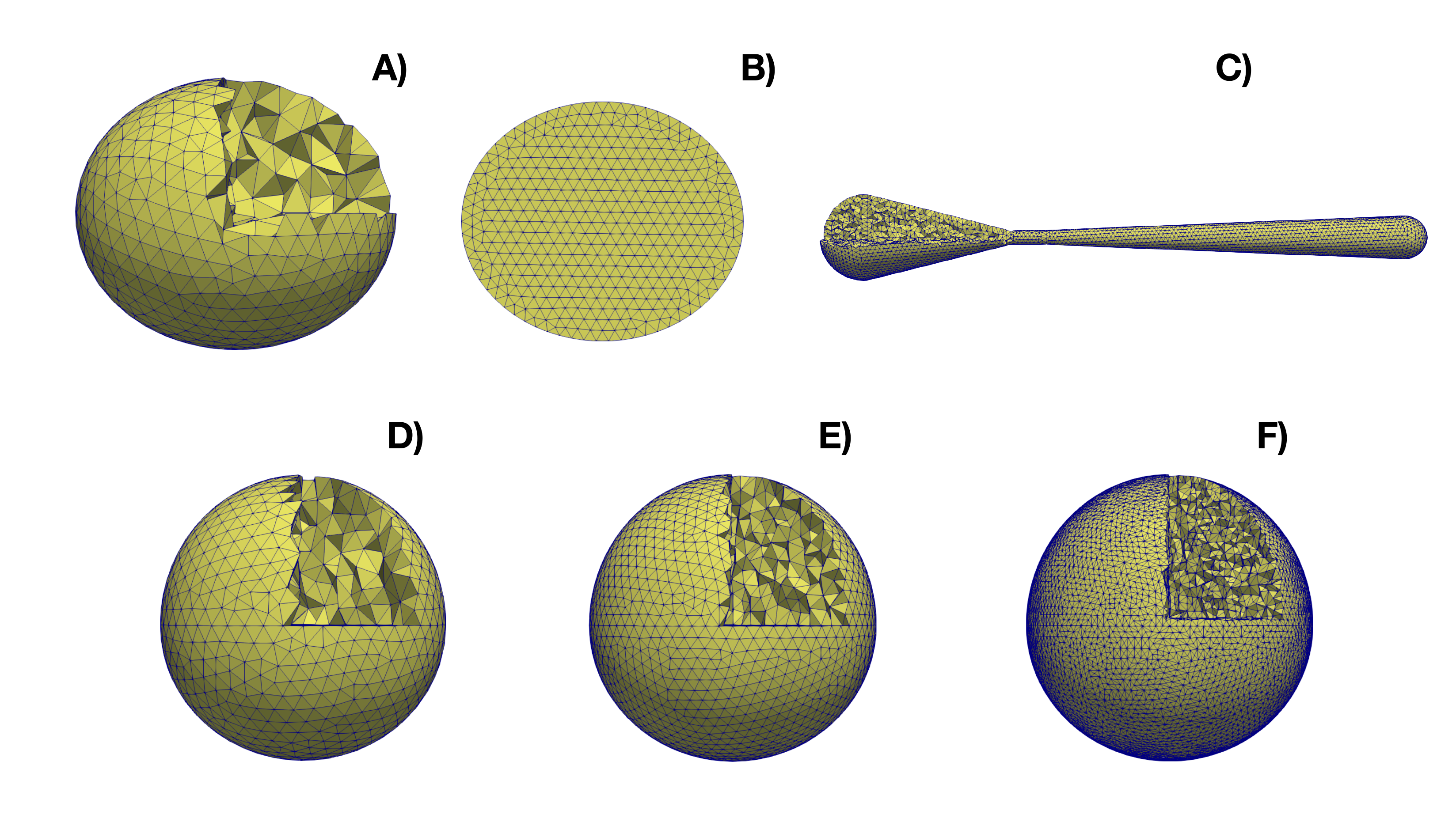}
    \caption{\re The meshes of the domains, with an opening on the bulk mesh. A) Mesh on the oblate spheroid, used in Figures \ref{fig:arclenghtsolution}-\ref{fig:3D-different-D_bulk}, \ref{fig:TBC-3D_and_2D}-\ref{fig:TBC}; B) Mesh on the ellipse of Figure \ref{fig:TBC-3D_and_2D}; C) Mesh of the elongated, non-convex geometry, used in the simulations in Figure \ref{fig:nonconvex}; 
    Figures D-F represent different refinement levels of the mesh of Figure \ref{fig:TBC}. See Table \ref{tab:discretisation} for more details. }
    \label{fig:meshes}
\end{figure}

\begin{figure}
    \centering
    \includegraphics[width=1\textwidth]{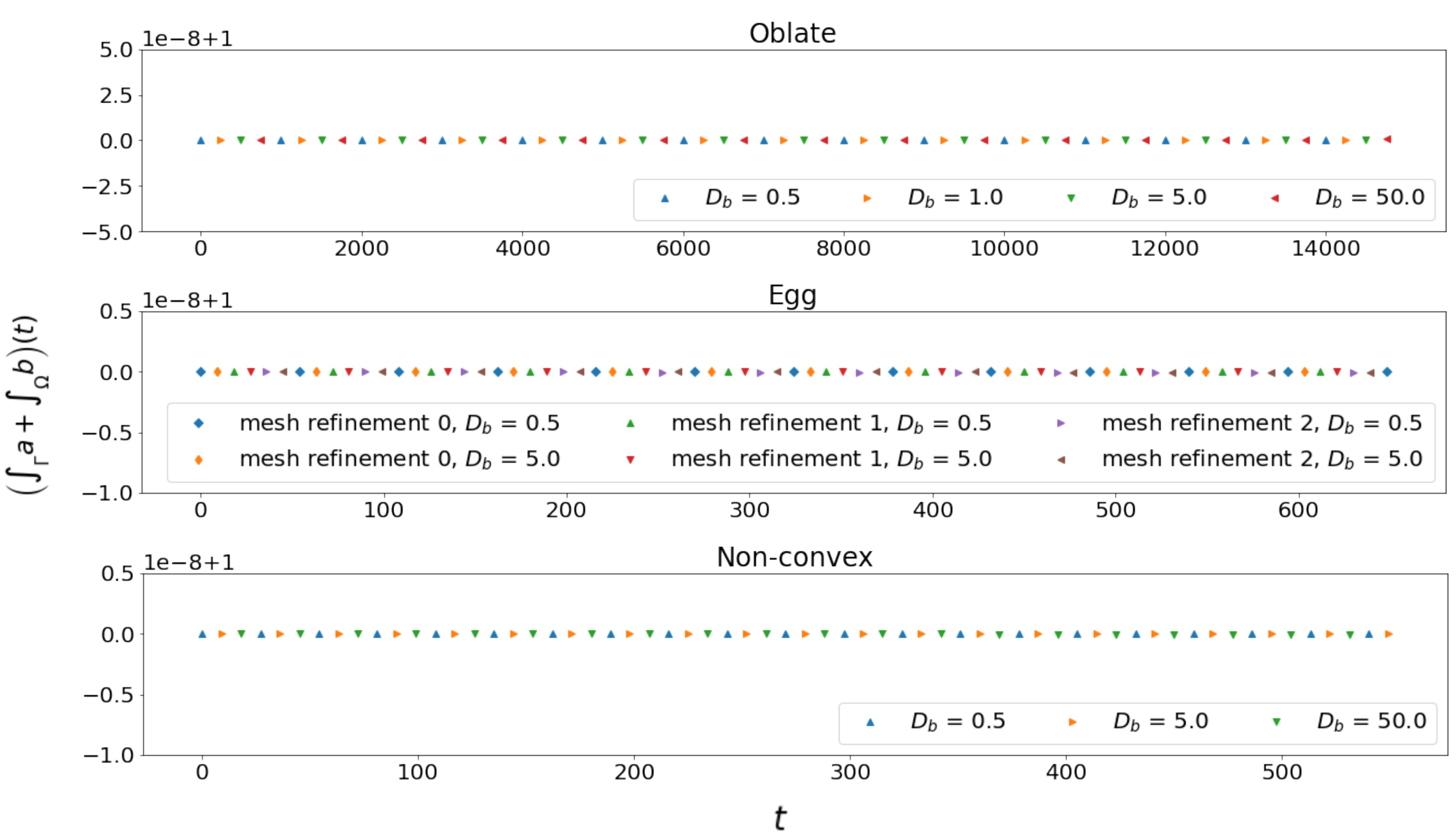}
    \caption{\re The numerical method conserves the total mass. In the figures we plot the rescaled total mass $\left(\int_\Gamma a + \int_\Omega b\right)/M_0$ of all the simulations of the BSWP model proposed in this manuscript. The figure on top refers to the simulations in Figure \ref{fig:3D-different-D_bulk}, the one in the middle to the simulations in Figure \ref{fig:TBC} for the three different mesh refinements (as shown in Fig. \ref{fig:meshes}), the figure on the bottom to the simulations in Figures \ref{fig:nonconvex} and \ref{fig:zoom_on_cones}.
    The maximal error, calculated as $\max_t \left(\int_\Gamma a + \int_\Omega b\right)(t)  - \min_t \left(\int_\Gamma a + \int_\Omega b \right)(t)$, is, respectively for each of the three figures, of the order $10^{-9}$, $10^{-10}$ and $10^{-11}$.}
    \label{fig:plot_masses}
\end{figure}

The parameters used throughout this manuscript are:
$D_a = 0.005$,
$k_0 = 0.1$,
$\gamma = 1.0$, 
$K = 0.3$,
$\beta = 1.0$. The parameter $D_b$ varies as indicated in the text. Unless otherwise stated, $M_0=|\Omega|$.
{\re For Figures \ref{fig:arclenghtsolution}-\ref{fig:L2norm-oblate-bulk-diffusion},
$a_{p,0}=1$, $\sigma_{x,0}^{-2}=\sigma_{y,0}^{-2}=\sigma_{z,0}^{-2}= 10$.
For Figures  \ref{fig:TBC-3D_and_2D} and  \ref{fig:TBC}, the initial condition \eqref{eq:TBC_ic} has parameters $a_{p,i}=1$, $\sigma_{v,i}^{-2}=10$ and $x_i=z_i=0$ for $v=x,y,z$ and $i=0,1$. In Figure  \ref{fig:TBC-3D_and_2D} we set $y_0=-y_1=0.85$, while, in Figure \ref{fig:TBC}, $y_0 = 1.05$ and $y_1=0.95$.
For Figure \ref{fig:nonconvex}, the initial condition \eqref{eq:ic_a} is prescribed with $a_{p,0}=1$, $\sigma_{y,0}^{-2}=15$,  $\sigma_{x,0}^{-2}=\sigma_{z,0}^{-2}=0$,  and $x_0=z_0=0$, $y_0=2.6$.
 As a last remark we want to stress out that, in this work, we always consider a nondimensional system and analyse it for certain parameter values within the range of wave-pinning dynamics. However our choices are still close to previous works. For instance, the kinetic parameters of the reaction function $f$ are very close to the ones in \cite{Mori2008, Cusseddu2019}. The parameter $K$ is reduced from 1 to $0.3$, which speeds up the positive feedback in activation. Such choice may be seen as a rescaling of the variables $a$ and $b$ and it is similar to \cite{Diegmiller2018,Miller2022}. Regarding the choice of the diffusion parameters, these are very closely related to the ones presented by Mori et al. \cite{Mori2008}, but with some adjustments. Indeed in \cite{Mori2008} $D_a = 0.1 \mu$m$^2/s$ and $D_b = 10 \mu$m$^2/s$ but they consider a cell diameter of 10$\mu$m which, directly extended to three-dimensions, would lead to a surface area of approximately 314 $\mu$m$^2$. In our cases, the surface area $\Gamma$ is between $11$ and $14$ (see Table \ref{tab:discretisation}). In dimensional units we may write $|\Gamma|\approx 12.5 X^2$, where $X$ is a unit dimension. 
Imposing $|\Gamma| = 314\mu$m$^2$ we find that $\mu$m$^2\approx0.04 X^2$ which gives us $D_a = 0.004 X^2/s$.
The bulk diffusion parameter $D_b$ is considered to be a couple of orders bigger than its surface counterpart \cite{Mori2008, Cusseddu2019, kamps2020optogenetic}. Therefore, since the aim of our work was to consider also the infinite bulk diffusion, we consider $D_b$ to be always of two orders of magnitude greater than $D_a$ (with the only exception in Figure \ref{fig:arclenghtsolution}, where the solution $a$ is reported also for $D_b = 0.05$, but polarisation fails).  }

\begin{table}\re
\footnotesize
\centering
\begin{tabular}{|l|l|c|c|c|c|c|c|} 
\hline
\multicolumn{2}{|l|}{Figure \ref{fig:meshes}} 
& {A)} 
& {B)}
& {C)}
& {D)}
& {E)}
& {F)}                                 \\ 
\hline
\multirow{5}{*}{\begin{turn}{90} Bulk $\Omega$ \end{turn} }     
& $|\Omega|$          
& 3.535574                                          & 2.666282    
& 2.087090 
& 4.166440 
& 4.178326  
& 4.178326                                         \\ 
\cline{2-8}
& vertices            
& 1587                                              & 447                               
& 10248                                             & 2465      
& 6500      
& 48479                                             \\ 
\cline{2-8}
& elements            
& 7140                                              & 826 
& 47937                                             & 11836     
& 33244     
& 265952                                            \\ 
\cline{2-8}
& $h_{max}$             
& 0.282133                                          & 0.110144                           
& 0.133535                                          & 0.263361  
& 0.180844  
& 0.128992                                          \\ 
\cline{2-8}
& $h_{min}$             
& 0.130494                                          & 0.068511 
& 0.057867                                          & 0.119002  
& 0.079174  
& 0.039587                                          \\ 
\hline
\multirow{5}{*}{\begin{turn}{90} Boundary $\Gamma$ \end{turn} } 
& $|\Gamma|$ 
& 11.286828                                         & 5.819239
& 13.870051                                         & 12.531459 
& 12.551124 
& 12.551124                                         \\ 
\cline{2-8}
& vertices 
& 815  
& 66
& 4749 
& 1050      
& 2238      
& 8946                                              \\ 
\cline{2-8}
& elements
& 1626                                              & 66
& 9494 
& 2096      
& 4472
& 17888                                             \\ 
\cline{2-8}
& $h_{max}$
& 0.282133 
& 0.088184
& 0.083116  
& 0.226042  
& 0.143022  
& 0.080545                                          \\ 
\cline{2-8}
& $h_{min}$
& 0.102122 
& 0.088150
& 0.041440
& 0.080562  
& 0.058842 
& 0.029241 
\\ 
\hline
\multicolumn{2}{|l|}{Time step}
& 0.01 
& 0.01 
& 0.01   
& 0.01  
& 0.01     
& 0.01                                               \\
\hline
\end{tabular}
\caption{\re Spatial and temporal discretisation. The table reports the number of vertices and elements of each mesh, together with information on the bulk, surface and mesh size. The values $h_{min}$ and $h_{max}$ indicate, respectively, the minimal and the maximal cell size in the mesh, which indicates the smallest and greatest distance between any two vertices of a mesh element.}\label{tab:discretisation}
\end{table}

 \bibliographystyle{elsarticle-num} 
 \bibliography{MS-arxiv}






\newpage
\begin{figure}
\centering
    \includegraphics[width = 1.1\textwidth]{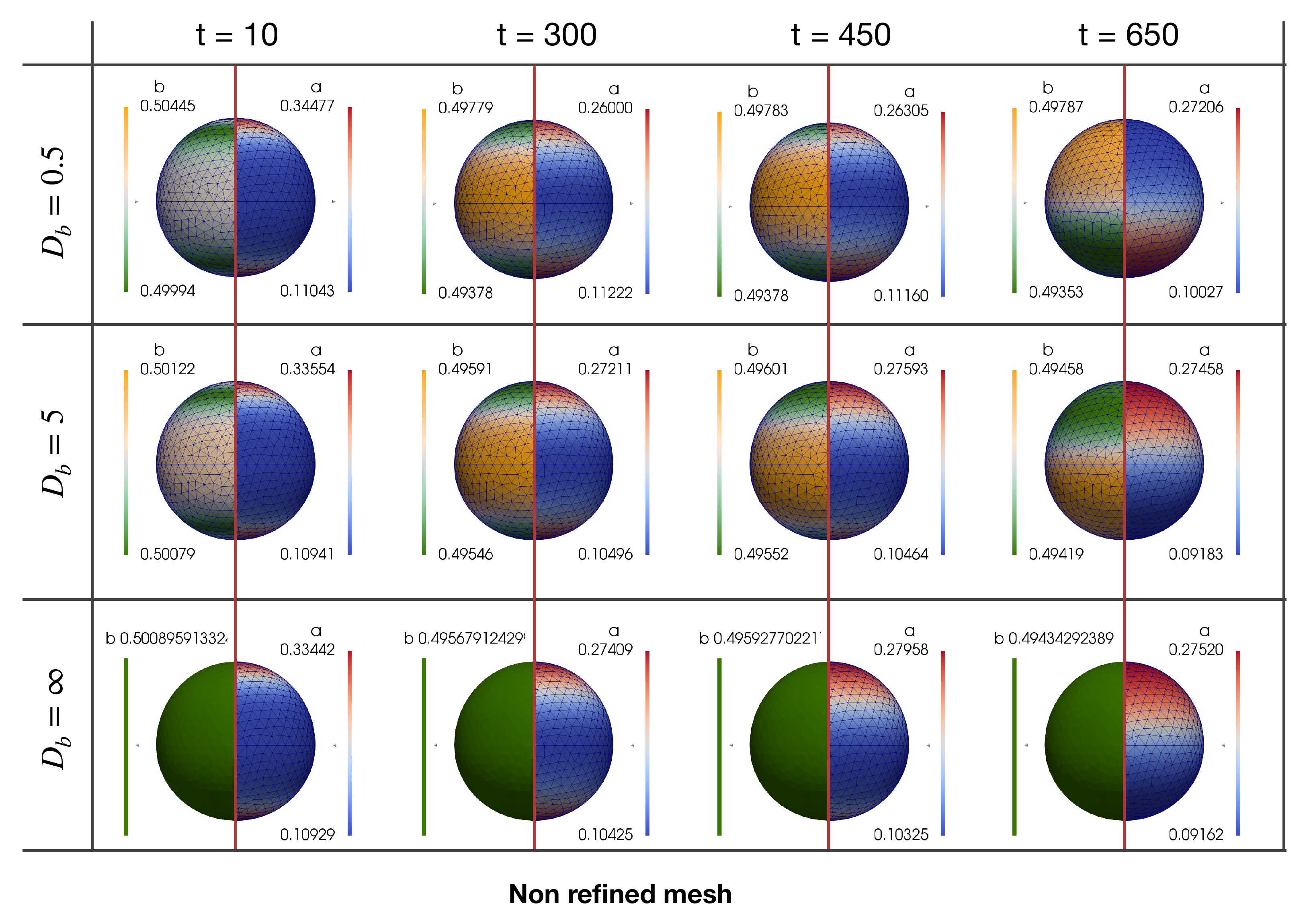}
    \captionsetup{labelformat=empty}
    \caption{\normalsize{\textbf{Supplementary figure. Cusseddu and Madzvamuse 2022.}\\ \textit{Patch competition}. The same results as in Figure \ref{fig:TBC} over the less refined mesh shown in Figure \ref{fig:meshes}D.}
    }
\end{figure}
\newpage
\begin{figure}
\centering
    \includegraphics[width = 1.1\textwidth]{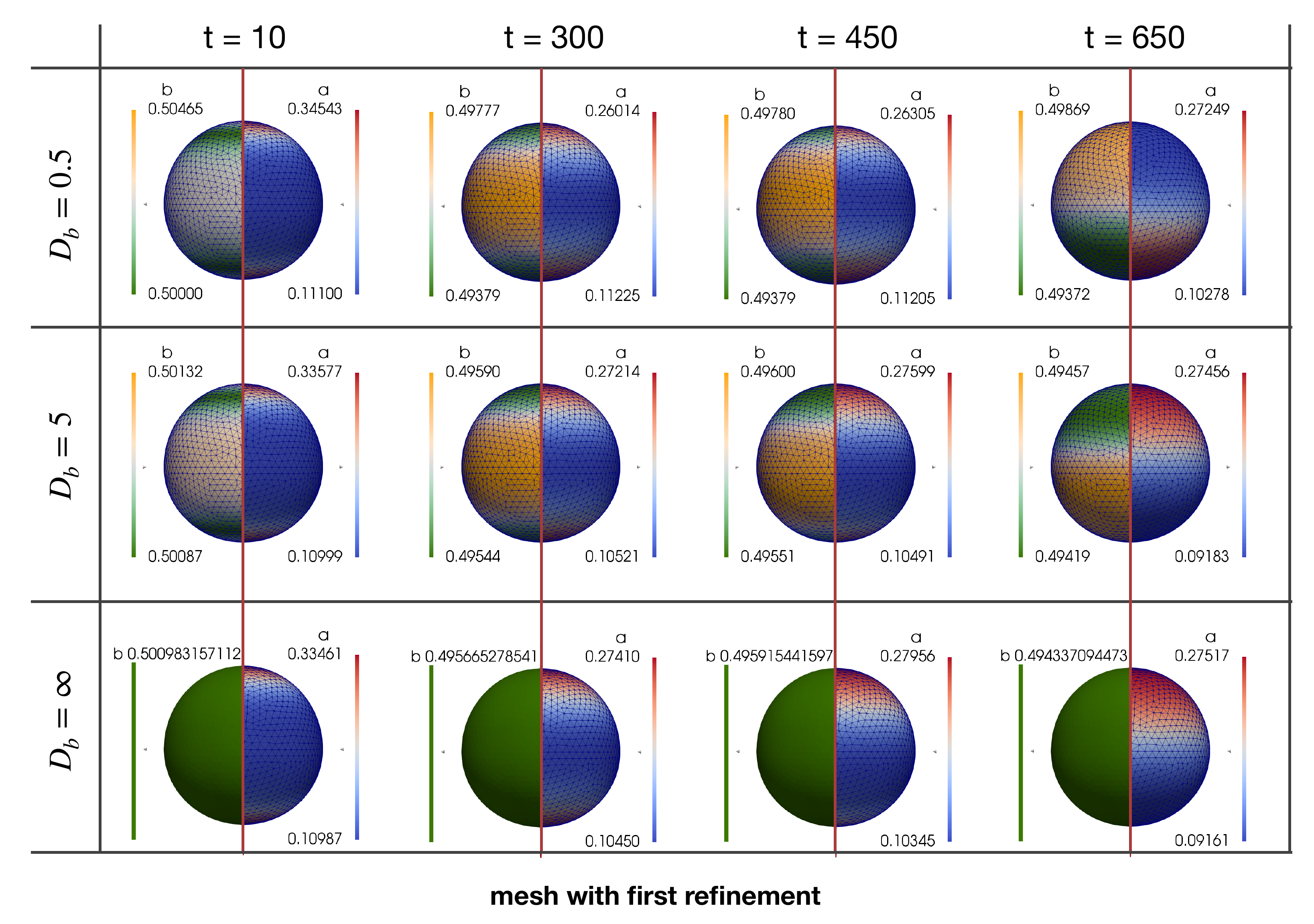}
    \captionsetup{labelformat=empty}
    \caption{\normalsize{\textbf{Supplementary figure. Cusseddu and Madzvamuse 2022.}\\ \textit{Patch competition}. The same results as in Figure \ref{fig:TBC} over the refined mesh shown in Figure \ref{fig:meshes}E.}
    }
\end{figure}
\end{document}